\documentclass[conference]{IEEEtran}

\IEEEoverridecommandlockouts
\usepackage{url}
\usepackage{cite}
\usepackage{amsmath,amssymb,amsfonts}
\usepackage{algorithmic}
\usepackage{graphicx}
\usepackage{textcomp}
\usepackage{booktabs}    
\usepackage{graphicx}    
\usepackage{array}       
\usepackage{float}
\usepackage{booktabs}  
\usepackage{caption}   
\usepackage{multirow}  
\usepackage{makecell}  
\usepackage{pifont}
\usepackage[table]{xcolor}  
\definecolor{mblue}{RGB}{0, 77, 128}
\definecolor{mblue}{RGB}{0, 77, 128}
\definecolor{mred}{RGB}{192,0, 0}
\definecolor{darkgreen}{rgb}{0.0, 0.5, 0.0}
\definecolor{mycolor_blue}{HTML}{E7EFFA}
\definecolor{mycolor_gray}{HTML}{ECECEC}
\newcommand{\cmark}{\textcolor{darkgreen}{\ding{51}}}  
\newcommand{\xmark}{\textcolor{red}{\ding{55}}}        

\def\BibTeX{{\rm B\kern-.05em{\sc i\kern-.025em b}\kern-.08em
    T\kern-.1667em\lower.7ex\hbox{E}\kern-.125emX}}
\begin{document}

\title{3DGS-VBench: A Comprehensive Video Quality Evaluation Benchmark for 3DGS Compression\\

\vspace{-0.5em}
 \author{
     Yuke Xing\textsuperscript{1},
     William Gordon\textsuperscript{2},
     Qi Yang\textsuperscript{3},
     Kaifa Yang\textsuperscript{1},
     Jiarui Wang\textsuperscript{1},
     Yiling Xu\textsuperscript{1}\\
     \textsuperscript{1}\{xingyuke-v, sekiroyyy, wangjiarui, yl.xu\}@sjtu.edu.cn, 
     \textsuperscript{2}williamg.research@gmail.com, 
     \textsuperscript{3}qiyang@umkc.edu \\
     1 Shanghai Jiao Tong University, Shanghai, China \\
     2 Basis Independent Silicon Valley, San Jose, CA, USA \\
     3 University of Missouri-Kansas City, Kansas City, USA \\
     
     \vspace{-3em}
 }
}



\maketitle

\begin{abstract}
3D Gaussian Splatting (3DGS) enables real-time novel view synthesis with high visual fidelity, but its substantial storage requirements hinder practical deployment, prompting state-of-the-art (SOTA) 3DGS methods to incorporate compression modules. However, these 3DGS generative compression techniques introduce unique distortions lacking systematic quality assessment research.
To this end, we establish \underline{\textit{3DGS-VBench}}, a large-scale \underline{V}ideo Quality Assessment (VQA) Dataset and \underline{Bench}mark with 660 compressed \underline{3DGS} models and video sequences generated from 11 scenes across 6 SOTA 3DGS compression algorithms with systematically designed parameter levels. With annotations from 50 participants, we obtained MOS scores with outlier removal and validated dataset reliability.
We benchmark 6 3DGS compression algorithms on storage efficiency and visual quality, and evaluate 15 quality assessment metrics across multiple paradigms. Our work enables specialized VQA model training for 3DGS, serving as a catalyst for compression and quality assessment research. The dataset is available at \url{https://github.com/YukeXing/3DGS-VBench}.
\end{abstract}


\section{Introduction}

\begin{figure*}[!t]
\vspace{-3mm}
    \centering
    \includegraphics[width=0.95\linewidth]{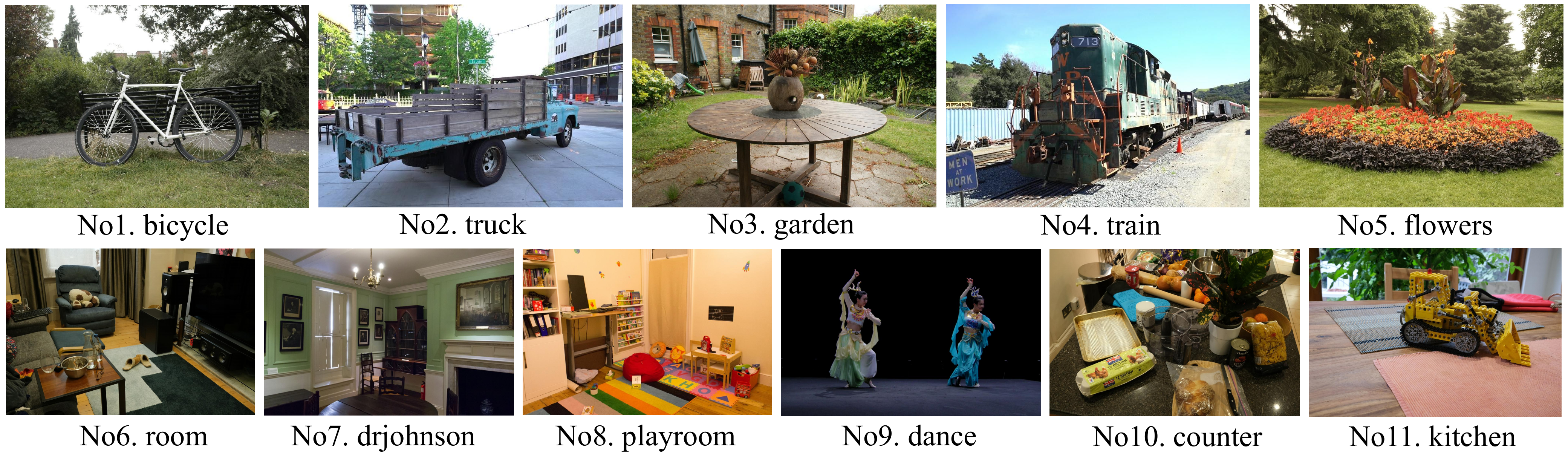}
    \caption{11 selected scenes in 3DGS-VBench: scenes 1-5 depict outdoor scenes, while scenes 6-11 depict indoor scenes.} 
    \label{figure:scene}
    \vspace{-5mm}
\end{figure*}

3D Gaussian Splatting (3DGS) \cite{3DGS} has emerged as a transformative approach for novel view synthesis (NVS), offering real-time rendering capabilities and visual fidelity compared to NeRF \cite{nerf}. However, its explicit representation demands substantial storage resources, severely hindering practical deployment, which prompting state-of-the-art (SOTA) 3DGS algorithms \cite{Scaffold-GS,HAC,Lightgaussian,CompGS,c3dgs,Compact-3DGS,eagles,liu2024compgs, yang2025hybridgs} to incorporate compression modules to balance visual quality and storage efficiency. These generative compression techniques introduce unique distortions that necessitate specialized quality assessment (QA) models. Thus, the scale and diversity of quality evaluation datasets for training and evaluation play a crucial role in developing accurate and robust 3DGS compression and QA models.

As we know, there have been 7 studies on NVS QA dataset, experiencing a transition from NeRF-oriented to 3DGS-centered evaluation. NeRF-QA \cite{NeRF-QA} and NeRF-VSQA \cite{NeRF-VSQA}, featuring 48 and 88 video samples respectively, while FFV \cite{FFV} established pairwise evaluation protocols using 220 samples. Systematic distortion analysis was advanced by ENeRF-QA \cite{ENeRF-QA}, which incorporated NeRF compression methodologies across 440 samples and characterized 9 distinct NeRF-related distortion categories. The emergence of 3DGS as a dominant NVS paradigm prompted corresponding shifts in evaluation priorities. Compression-related QA was explored by GSC-QA \cite{GSC-QA} using 120 samples to analyze their specific 3DGS compression approach, while GS-QA \cite{GS-QA} conducted comparative analysis between 3DGS and NeRF methodologies across 64 samples. Both video and image evaluation domains were addressed by NVS-QA \cite{NVS-QA}, incorporating 65 samples for each modality.

However, existing 3DGS QA datasets exhibit some critical limitations hindering comprehensive evaluation. (1) \textbf{Limited distortion types.} Despite compression becoming essential for 3DGS practical deployment due to substantial storage demands, most current datasets predominantly fail to incorporate diverse generative compression strategies. This oversight results in insufficient samples with diverse compression-induced distortion effects for effective metric training. (2) \textbf{Limited dataset scale.} Due to substantial costs of  subjective experiments, existing datasets contain fewer than 100 samples with the largest not exceeding 500. More critically, none of the existing datasets comprise over 100 authentic real-world scene samples, severely insufficient for training robust deep learning-based QA models that simulate human perception standards.

To address these limitations, we present \textbf{3DGS-VBench}, a large-scale dataset and benchmark for 3DGS video quality evaluation, containing 660 3DGS models and corresponding video sequences. The contents consist of  11 real-world scenes selected from mainstream multi-view datasets \cite{Mip-NeRF-360, Tanks-Temples, Deep-Blending}. 6 representative mainstream 3DGS compression algorithms \cite{Compact-3DGS,CompGS,c3dgs,Scaffold-GS,Lightgaussian,HAC} with systematically designed multi-level compression parameters are considered, leading to diverse distortion and large sample scale. After rendering the 3DGS samples into videos and conducting subjective experiments, we collect 10k expert ratings to calculate mean opinion scores (MOS).
Based on 3DGS-VBench, we benchmark the
6 representative 3DGS models across storage efficiency and visual quality dimensions.
Besides, leveraging the substantial scale of our dataset, we conduct a comprehensive QA metrics evaluation benchmark with diverse IQA/VQA metrics across multiple evaluation paradigms on 3DGS content. This benchmark reveals the limitations of existing QA algorithms, as well as the characteristics of
SOTA 3DGS compression models in terms of storage efficiency and visual quality.
The main contributions of this paper are summarized as follows:
\begin{itemize}
\item We establish \textbf{3DGS-VBench}, a large-scale dataset comprising 660 3DGS-rendered videos with diverse 3DGS compression distortions annotated with MOS scores.
\item Based on 3DGS-VBench, we {benchmark and highlight the characteristics of SOTA 3DGS compression methods from two performance dimensions}: storage efficiency and visual quality.
\item We evaluate and benchmark 15 QA metrics on 3DGS-VBench. Thorough analysis is provided, we report the weakness of  current 3DGS QA study.  
\end{itemize}

\section{Dataset Construction}

\subsection{Source Content Selection}

To achieve comprehensive quality assessment with diverse visual characteristics, we select 11 real-world scenes from 4 well-established multiview datasets as shown in Fig. \ref{figure:scene}. Six scenes are picked from Mip-NeRF360 \cite{Mip-NeRF-360}, including three outdoor scenes: bicycle (1237 × 822), flowers (1256 × 828), and garden (1297 × 840), and three indoor scenes: counter (1558 × 1038), kitchen (1558 × 1039), and room (1557 × 1038). Two outdoor scenes are selected from Tanks \& Temples \cite{Tanks-Temples}: train (980 × 545) and truck (979 × 546). Two indoor scenes are selected from Deep Blending \cite{Deep-Blending}: playroom (1264 × 832) and drjohnson (1332 × 876). One human figure scene Dance\_Dunhuang\_Pair (dance) with resolution 1600 × 876 is selected from PKU-DyMVHumans \cite{pku}.

\subsection{Camera Parameter Setup for Video Rendering}

Existing classical 3DGS real scene multi-view image datasets \cite{Mip-NeRF-360, Tanks-Temples,Deep-Blending} consist of discrete camera viewpoints, divided into training and testing sets. The training set optimizes 3DGS models, while the testing set provides reference images for quality assessment of rendered views from the reconstructed scene. However, video rendering requires continuous camera trajectories, which are absent in these multi-view datasets containing only sparse, discrete viewpoints. This necessitates generating dense, continuous camera paths for video evaluation, which we name it as ``Val'' viewpoint set.

To create the ``Val'' set, first, we train 3DGS models to obtain pointcloud representations of selected scenes. Then, we import these pointclouds into Blender where we establish virtual camera setups with carefully calibrated parameters including focal length, position, orientation, and resolution settings. Through iterative adjustment, we design 600 continuous viewpoints for each scene, maintaining resolution consistency with original datasets. These viewpoints form a $360^{\circ}$ orbital trajectory for smooth video sequence generation.

\subsection{3DGS Model Selection \& Compression Parameter Design}


To systematically investigate visual distortion effects in 3DGS compression, we select 6 representative mainstream algorithms and analyze their compression strategies and key parameters. For each compression parameter, we establish multiple \textbf{Compression Levels (CL)} to control distortion intensity. For each 3DGS training process, we vary only one compression parameter with different CL while other compression parameters maintain the default optimal values. This methodology can generate 3DGS models with diverse \textbf{distortion levels (DL)} and types, establishing comprehensively study for visual distortion effects introduced by different 3DGS compression strategies. The selected 3DGS algorithms and detailed CL configurations are presented as follows:

\paragraph{\textbf{Compact-3DGS}~\cite{Compact-3DGS}} achieves compression through learnable volume masking (lambda parameter), residual vector quantization (RVQ) for geometry encoding (codebook size and RVQ depth), and hash grid-based neural fields replacing spherical harmonics (SH) (hashmap parameter).  As illustrated in table \ref{tab:Compact-3DGS}, with 3 key compression parameters each having 5 CL where {CL5 represents the default optimal configuration}, 
\textbf{we have \boldmath 3 compression parameters $\times$ 4 distorted CL = 12 distorted samples, and 1 default sample with all optimal parameters, resulting in 13 DL per scene}.

\begin{table}[H]
\vspace{-1mm}
  \caption{CL for compression parameters in \textbf{Compact-3DGS}.}
  \renewcommand{\arraystretch}{1}
  \label{tab:Compact-3DGS}
  \centering
  \vspace{-1.5mm}
  \resizebox{0.48\textwidth}{!}{
  \renewcommand{\arraystretch}{1}
  \begin{tabular}{c!{\vrule}ccccc}
    \toprule
    \textbf{Parameter} & \textbf{CL1} & \textbf{CL2} & \textbf{CL3} & \textbf{CL4} & \textbf{CL5 (default)}
    \\
    \midrule
    \textbf{hashmap} & $2^1$ & $2^2$ & $2^7$ & $2^{12}$ & $2^{19}$ 
    \\
    \textbf{(codebook, rvq\_depth)} & ($2^3$, 1) & ($2^4$, 1) & ($2^6$, 1) & ($2^3$, 6) & ($2^6$, 6)
    \\
    \textbf{lambda} & 0.014 & 0.012 & 0.010 & 0.006 & 0.0005
    \\
    \bottomrule  
    
  \end{tabular}}
\vspace{-1mm}
\end{table}

\paragraph{\textbf{CompGS}~\cite{CompGS}} employs $K$-means vector quantization of Gaussian parameters, controlled by geometry codebook size (g-size), color codebook size (c-size), and opacity regularization parameter (reg). As shown in table \ref{tab:CompGS}, with 3 key compression parameters and 5 CL each, 
we generate \boldmath $3  \times 4 + 1  = 13$ \textbf{DL} per scene.

\begin{table}[H]
\vspace{-1mm}
  \caption{CL for compression parameters in \textbf{CompGS}.}
  \renewcommand{\arraystretch}{1}
  \label{tab:CompGS}
  \centering
  \vspace{-1.5mm}
  \resizebox{0.48\textwidth}{!}{
  \begin{tabular}{c!{\vrule}ccccc}
    \toprule
    \textbf{Parameter} & \textbf{CL1} & \textbf{CL2} & \textbf{CL3} & \textbf{CL4} & \textbf{CL5 (default)}
    \\
    \midrule
    \textbf{g-size} & 1 & 2 & $2^3$ & $2^5$ & $2^{12}$ 
    \\
    \textbf{c-size} & 1 & 2 & $2^2$ & $2^3$ & $2^{12}$
    \\
    \textbf{reg} & 4$\times$1e-6 & 3$\times$1e-6 & 2$\times$1e-6 & 1$\times$1e-6 & 1$\times$1e-7
    \\
    \bottomrule  
    
  \end{tabular}}
\vspace{-1mm}
\end{table}

 \paragraph{\textbf{c3dgs}~\cite{c3dgs}} uses $K$-means to construct color and geometry codebooks with entropy encoding. Compression is controlled by codebook size and importance thresholds (c-size/c-include for color, g-size/g-include for geometry). As illustrated in table \ref{tab:c3dgs}, with 2 key compression parameters each having 5 CL, 
we generate {\boldmath $2  \times 4  + 1  = 9$ DL per scene}.

\begin{table}[H]
\vspace{-1mm}
  \caption{CL for compression parameters in \textbf{c3dgs}.}
  \renewcommand{\arraystretch}{1.2}
  \label{tab:c3dgs}
  \centering
  \vspace{-1.5mm}
  \resizebox{0.48\textwidth}{!}{
  \renewcommand{\arraystretch}{1}
  \begin{tabular}{c!{\vrule}ccccc}
    \toprule
    \textbf{Parameter} & \textbf{CL1} & \textbf{CL2} & \textbf{CL3} & \textbf{CL4} & \textbf{CL5 (default)}
    \\
    \midrule
    \textbf{(c-size, c-include)} & (1, 0.6) & (2, 0.6) & ($2^2$, 0.6) & ($2^5$, 0.6) & ($2^{12}$, 0.6$\times$1e-6) 
    \\
    \textbf{(g-size, g-include)} & (1, 0.3) & ($2^3$, 0.3) & ($2^5$, 0.3) & ($2^8$, 0.3) & ($2^{12}$, 0.3$\times$1e-5)
    \\
    \bottomrule   
    
  \end{tabular}}
\vspace{-1mm}
\end{table}

\paragraph{\textbf{LightGaussian}~\cite{Lightgaussian}} compresses through pruning low-significance Gaussians, distilling SH coefficients to lower degrees and quantizing them via vector quantization. Compression is controlled by pruning ratio (prune) and SH quantization parameters (c-ratio/c-size). As shown in table \ref{tab:LightGaussian}, 
we generate {\boldmath $2  \times 4 + 1  = 9$ DL per scene}.

\begin{table}[H]
\vspace{-1mm}
  \caption{CL for parameters in \textbf{LightGaussian}.}
  \renewcommand{\arraystretch}{1}
  \label{tab:LightGaussian}
  \centering
  \vspace{-1.5mm}
  \resizebox{0.48\textwidth}{!}{
  \begin{tabular}{c!{\vrule}ccccc}
    \toprule
    \textbf{Parameter} & \textbf{CL1} & \textbf{CL2} & \textbf{CL3} & \textbf{CL4} & \textbf{CL5 (default)}
    \\
    \midrule
    \textbf{prune} & 0.97 & 0.95 & 0.90 & 0.85 & 0.66 
    \\
    \textbf{(c-ratio, c-size)} & (1, 2) & (1, $2^2$) & (1, $2^3$) & (1, $2^5$) & (1, $2^{13}$)
    \\
    \bottomrule   
    
  \end{tabular}}
\vspace{-1mm}
\end{table}

\paragraph{\textbf{Scaffold-GS}~\cite{Scaffold-GS}} uses anchor-based hierarchical sampling, with compression controlled by voxel size parameter (vsize). As shown in Table \ref{tab:Scaffold-GS}, with $1$ compression parameter corresponding to $8$ CL, we generate \boldmath$8$ DL per scene.

\begin{table}[H]
\vspace{-2mm}
  \caption{CL for compression parameter in \textbf{Scaffold-GS}.}
  \renewcommand{\arraystretch}{1}
  \label{tab:Scaffold-GS}
  \centering
  \vspace{-1.5mm}
  \resizebox{0.48\textwidth}{!}{
  \begin{tabular}{c!{\vrule}cccccccc}
    \toprule
    \textbf{Param} & \textbf{CL1} & \textbf{CL2} & \textbf{CL3} & \textbf{CL4} & \textbf{CL5} & \textbf{CL6} & \textbf{CL7} & \textbf{CL8 (default)}
    \\
    \midrule
    \textbf{vsize} & 0.350 & 0.250 & 0.200 & 0.160 & 0.120 & 0.080 & 0.050 & 0.001
    \\
    \bottomrule  
  \end{tabular}}
\vspace{-1mm}
\end{table}

\paragraph{\textbf{HAC}~\cite{HAC}} 
besides using anchor-based hierarchical sampling, they employs context-aware compression with adaptive quantization controlled by the lambda parameter to balance rate-distortion performance. As shown in Table \ref{tab:HAC}, with $1$ compression parameter corresponding to $8$ CL, we generate \boldmath$8$ DL per scene.

\begin{table}[H]
\vspace{-2mm}
  \caption{CL for compression parameter in \textbf{HAC}.}
  \renewcommand{\arraystretch}{1}
  \label{tab:HAC}
  \centering
  \vspace{-1.5mm}
  \resizebox{0.48\textwidth}{!}{
  \begin{tabular}{c!{\vrule}cccccccc}
    \toprule
    \textbf{Param} & \textbf{CL1} & \textbf{CL2} & \textbf{CL3} & \textbf{CL4} & \textbf{CL5} & \textbf{CL6} & \textbf{CL7} & \textbf{CL8 (default)}
    \\
    \midrule
    \textbf{lmbda} & 0.800 & 0.600 & 0.400 & 0.300 & 0.200 & 0.120 & 0.060 & 0.004
    \\
    \bottomrule  
    
  \end{tabular}}
\vspace{-1mm}
\end{table}

\textbf{In all, we trained \boldmath 11 scenes $\times$ (13 + 13 + 9 + 9 + 8 + 8) model settings = 660 3DGS reconstruction models}.

\subsection{Subjective Experiment and Data Processing}
For subjective QA, we convert each 3DGS reconstruction into Processed Video Sequences (PVS) by rendering 600 frames that orbit each scene with uniform angular spacing established in  our "Val" camera trajectories. The frames are encoded into 20-second PVS using FFMPEG with libx265 codec at 30 fps and a constant rate factor of 10.

For the quality annotation, we use an 11-level impairment scale proposed by ITU-TP.910 \cite{ITU1(siti)}. The experiment was carried out using a 27-inch AOC Q2790PQ monitor in an indoor laboratory environment under standard lighting conditions. The videos are displayed using an interface designed with Python Tkinter. To prevent visual fatigue caused by excessively experiment time, 660 PVSs are randomly divided into 8 smaller groups. Finally, {we obtain a total of 9,900 human annotations (15 annotators × 660 videos)}.

We follow the suggestions recommended by ITU to conduct the outlier detection and subject rejection. The score rejection rate is 2\%.
At last, a total of {660 MOSs} are obtained.

\section{Dataset Validation}

\begin{table*}[t]
\vspace{-3mm}
\caption{Comprehensive Quantitative Comparison Benchmark for 6 representative 3DGS compression methods in 3 Real-world Multi-view images Datasets. The best results are marked in {\colorbox{red!20}{RED}} and the second-best in {\colorbox{mycolor_blue!200}{BLUE}}. }  
\renewcommand{\arraystretch}{1}
\label{tab:bench}
\centering
\vspace{-1.5mm}
\resizebox{1\textwidth}{!}{
\renewcommand{\arraystretch}{1.2}
  \begin{tabular}{l||ccccc|ccccc|ccccc}
    \toprule
    \multicolumn{1}{l}{\textbf{Dataset}} &  
    \multicolumn{5}{c}{\textbf{Mip-NeRF360}\cite{Mip-NeRF-360}} &
    \multicolumn{5}{c}{\textbf{Tanks\&Temples}\cite{Tanks-Temples}} &  
    \multicolumn{5}{c}{\textbf{Deep Blending}\cite{Deep-Blending}} \\
    \cmidrule(lr){2-6} \cmidrule(lr){7-11} \cmidrule(lr){12-16} 
    \textbf{Methods / Metrics} & \textbf{PSNR} $\uparrow$ & \textbf{SSIM}$\uparrow$ & \textbf{LPIPS}$\downarrow$ & \textbf{MOS}$\uparrow$ & \textbf{Size(MB)}$\downarrow$ & \textbf{PSNR}$\uparrow$ & \textbf{SSIM}$\uparrow$ & \textbf{LPIPS}$\downarrow$ & \textbf{MOS}$\uparrow$ & \textbf{Size(MB)}$\downarrow$ & \textbf{PSNR}$\uparrow$ & \textbf{SSIM}$\uparrow$ & \textbf{LPIPS}$\downarrow$ & \textbf{MOS}$\uparrow$ & \textbf{Size(MB)}$\downarrow$ \\
    \midrule  
    \rowcolor{gray!20} 3DGS (origin) \cite{3DGS}& 27.6655 & 0.9141 & 0.1267 & - &795.1323 & 23.7579 & 0.89339 & 0.09559 & - & 434.3680 & 29.6219 & 0.9269 & 0.1016& -
& 665.9170

\\
    Compact-3DGS  \cite{Compact-3DGS} & 27.0677& 0.9006& 0.1494& 8.1186& 48.7238& 23.3552& 0.8803& 0.1139& 6.9000& 39.7017& 29.7882& 0.9280& 0.1016& \colorbox{mycolor_blue!200}{8.6786}& 43.3126
\\
    c3dgs \cite{c3dgs} & 27.3182& 0.9067& 0.1374& 8.2960& 28.6562& 23.6088& 0.8880& 0.1003& 7.5619& 17.6700 & 29.5096& 0.9250& \colorbox{mycolor_blue!200}{0.0968}& \colorbox{mycolor_blue!200}{8.6786}& 23.8754
\\
    LightGaussian \cite{Lightgaussian} & 25.4406& 0.8582& 0.2009& 8.1333& 52.0151& 22.6449& 0.8565& 0.1637& 7.4190& 28.6682& 25.9883& 0.8447& 0.1950& 7.9333& 43.3730
\\
    CompGS \cite{CompGS}& 27.1742& 0.9042& 0.1468& 8.3484& \colorbox{mycolor_blue!200}{21.9188}& 23.2633& 0.8845& 0.1132& 7.5524& \colorbox{red!20}{13.6345}& 30.0751& 0.9338& 0.0991& \colorbox{red!20}{9.0238}& \colorbox{mycolor_blue!200}{14.7798}\\
    Scaffold-GS \cite{Scaffold-GS}  & \colorbox{red!20}{27.8406} & 
\colorbox{red!20}{0.9149}& \colorbox{red!20}{0.1285}& \colorbox{red!20}{8.8233}& 166.8277& \colorbox{mycolor_blue!200}{24.2154}& \colorbox{red!20}{0.9033}& \colorbox{red!20}{0.0868}& \colorbox{red!20}{8.5595}& 147.3997& \colorbox{mycolor_blue!200}{30.2592}& \colorbox{red!20}{0.9349}& \colorbox{red!20}{0.0933}& 8.7308& 111.8945
\\
    HAC \cite{HAC} & \colorbox{mycolor_blue!200}{27.5911} & \colorbox{mycolor_blue!200}{0.9112}& \colorbox{mycolor_blue!200}{0.1364}& \colorbox{mycolor_blue!200}{8.7413}& \colorbox{red!20}{15.8625}& \colorbox{red!20}{24.3093} & \colorbox{mycolor_blue!200}{0.9014}& \colorbox{mycolor_blue!200}{0.0928}& \colorbox{mycolor_blue!200}{8.1786}& \colorbox{mycolor_blue!200}{14.5386}& \colorbox{red!20}{30.3207}& \colorbox{mycolor_blue!200}{0.9342}& 0.0975& 8.8214& \colorbox{red!20}{7.7961}\\
    \bottomrule
  \end{tabular}
}
\label{benchmark}
\vspace{-3mm}
\end{table*}

\subsection{Diversity of Source Content}

To assess the diversity of selected source content, we analyzed the spatial information (SI) \cite{ITU1(siti)} and temporal information (TI) \cite{ITU1(siti)} characteristics of our 11 scenes, as presented in Fig. \ref{fig:SITI} (a), where the labels match the scene numbers in Fig. \ref{figure:scene}. The scattered distribution pattern demonstrates comprehensive coverage across both spatial and temporal complexity dimensions, indicating effective diversity in scene characteristics. Thus, our dataset provides a robust foundation for evaluating 3DGS across diverse visual content types.

\subsection{Analysis of MOS}
\begin{figure}[H]
    \centering
    \vspace{-1mm}
    \includegraphics[width=0.5\textwidth]{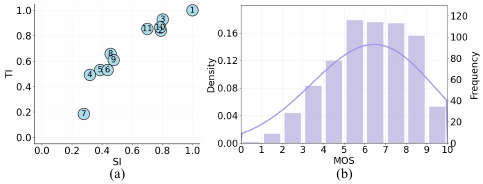}
    \vspace{-6mm}
    \caption{(a) SI vs. TI (b) MOS Distribution} 
    \label{fig:SITI}
    \vspace{-2mm}
\end{figure}

To verify the reasonability of MOS scores, Fig. \ref{fig:SITI} (b) shows the MOS distribution covers the full 0-10 range with an approximately Gaussian distribution centered at 6. The dataset maintains adequate representation across all quality segments, including severely degraded content (0-2), providing essential training diversity for robust QA model development.

\subsection{Analysis of CL Design}

To validate our MOS scores and compression level settings, we examine the MOS-storage relationship using LightGaussian as example and evaluating key compression parameters across five designed CL in Table \ref{tab:LightGaussian} on four representative scenes: \textit{truck} (Tanks \& Temples), \textit{drjohnson} (Deep Blending), \textit{flowers} and \textit{room} (Mip-NeRF360).

\begin{figure}[H]
    \centering
    \vspace{-1mm}
    \includegraphics[width=0.5\textwidth]{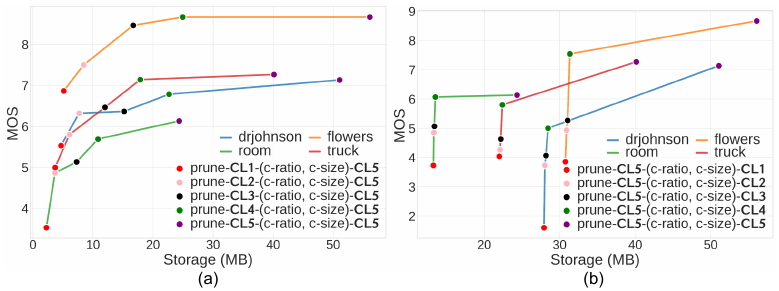}
    \vspace{-6mm}
    \caption{MOS vs. Storage (MB) for LightGaussian with diverse CL settings. (a) 5 CL for parameter \textit{prune}. (b) 5 CL for parameters \textit{(c-ratio, c-size)}.} 
    \label{fig:mos-storage}
    \vspace{-2mm}
\end{figure}

The results for geometry parameter \textbf{\textit{prune}} are shown in Fig. \ref{fig:mos-storage} (a), while Fig. \ref{fig:mos-storage} (b) shows results for color compression parameters \textbf{\textit{(c-ratio, c-size)}}. The clear monotonic relationships between storage size and MOS scores validate our scoring system and CL design, as larger models should inherently store more information and achieve higher visual quality.

Besides, the results reveal distinct compression behaviors: In Fig. \ref{fig:mos-storage} (a), \textbf{\textit{prune}} achieves significant storage reduction with minimal quality loss (CL3-CL5), while in Fig. \ref{fig:mos-storage} (b), \textbf{\textit{(c-ratio, c-size)}} cause substantial quality degradation with limited storage savings (CL1-CL4). This demonstrates that for LightGaussian, \textbf{\textit{prune}} provides superior quality-storage trade-offs for practical deployment.

\section{Benchmark for Quality Assessment Metrics}

To evaluate existing QA methods on compressed 3DGS content, we conduct correlation analysis between objective metrics and MOS scores using Spearman Rank-order Correlation Coefficient (SRCC), Pearson Linear Correlation Coefficient (PLCC), and Kendall Rank Correlation Coefficient (KRCC). 

\begin{table}[h]
\caption{Performance benchmark on 3DGS-VBench.  $\heartsuit$ various IQA models, $\spadesuit$ deep-learning-based VQA models.}
\renewcommand{\arraystretch}{0.8}
\label{tab:metric}
\centering
\resizebox{0.48\textwidth}{!}{
  \begin{tabular}{l||c|ccc}
    \toprule
    \textbf{Metrics} & \textbf{Ref} & \textbf{SRCC} & \textbf{PLCC} & \textbf{KRCC} \\
    \midrule
    $\heartsuit$ PSNR (TIP 2004) \cite{PSNRSSIM} & \cmark & 0.5022 & 0.4976 & 0.3560 \\
    $\heartsuit$ SSIM (TIP 2004) \cite{PSNRSSIM}& \cmark & 0.5108 & 0.4758 & 0.3684 \\
    $\heartsuit$ LPIPS (CVPR 2018) \cite{LPIPS}& \cmark & 0.5106 & 0.4581 & 0.3619 \\
    $\heartsuit$ DISTS (PAMI 2022) \cite{DISTS} & \cmark & 0.7317 & 0.7146 & 0.5269 \\
    $\heartsuit$ VIF (TIP 2006) \cite{VIF}& \cmark & 0.4220 & 0.4252 & 0.2917 \\
    $\heartsuit$ FSIM (TIP 2011) \cite{FSIM}& \cmark & 0.5866 & 0.5815 & 0.4311 \\
    $\heartsuit$ IW-SSIM (TIP 2011) \cite{IW-SSIM}& \cmark & 0.6309 & 0.5962 & 0.4488 \\
    $\heartsuit$ MS-SSIM (SSC 2003) \cite{MS-SSIM}& \cmark & 0.5028 & 0.4700 & 0.3602 \\
    $\heartsuit$ CLIP-IQA (AI 2023) \cite{CLIP-IQA}& \xmark & 0.3913 & 0.2738 & 0.3216 \\
    $\heartsuit$ BRISQUE (TIP 2012) \cite{BRISQUE}& \xmark & 0.2379 & 0.1819 & 0.1749 \\

    \midrule
    $\spadesuit$ DOVER (ICCV 2023) \cite{DOVER} & \xmark & 0.9409 &  0.9308 & 0.7901 \\
    $\spadesuit$ FAST-VQA (ECCV 2022) \cite{FAST-VQA}& \xmark & 0.9314 & 0.9255 & 0.7753 \\
    $\spadesuit$ simpleVQA (ICM 2022) \cite{simpleVQA}& \xmark & 0.9350 & 0.7813 & 0.9314 \\
    $\spadesuit$ VSFA (ICM 2019) \cite{VSFA}& \xmark & 0.9392 & 0.9345 & 0.7908 \\
    $\spadesuit$ Q-Align (PAMI 2023) \cite{Q-Align}& \xmark & 0.8485 & 0.6489 & 0.8126 \\
    \bottomrule
  \end{tabular}
}
\label{bench-metric}
\vspace{-1mm}
\end{table}

We test 15 objective metric: i) full-reference image metrics, including classic metrics: PSNR, SSIM \cite{PSNRSSIM}, MS-SSIM \cite{MS-SSIM}, IW-SSIM \cite{IW-SSIM}, VIF \cite{VIF} and FSIM \cite{ FSIM}, and deep-learning based metrics: LPIPS \cite{LPIPS} and DISTS \cite{DISTS};
ii) no-reference image metrics, including BRISQUE \cite{BRISQUE}, and LLM-pretraining metric CLIP-IQA \cite{CLIP-IQA}; 
iii) deep learning-based no-reference video metrics, including
DOVER \cite{DOVER}, FAST-VQA \cite{FAST-VQA} simpleVQA, and Q-Align \cite{Q-Align}, as there's often no reference videos in typical multi-view image datasets.

As shown in Table \ref{tab:metric}, deep learning-based VQA models substantially outperform traditional approaches, with DOVER achieving highest correlation (SRCC=0.9409, PLCC=0.9308, KRCC=0.7901). SimpleVQA, VSFA, and FAST-VQA also show strong performance (about 0.93 SRCC). Among traditional metrics, DISTS performs best (SRCC=0.7317), while no-reference methods show poor performance (BRISQUE SRCC=0.2379). Since current 3DGS generation uses L1 loss and SSIM supervision, we think using better metrics as loss functions \cite{MPED} could improve generation results.

\section{Compression Benchmark for 3DGS Models}

To investigate current 3DGS compression performance across visual quality and storage efficiency, we comprehensively compared 6 representative methods on 3 datasets using default optimal parameters, establishing the first standard benchmark for 3DGS compression algorithms.

As shown in Table \ref{tab:bench}, anchor-based method Scaffold-GS achieves superior visual quality exceeding original 3DGS but limited compression ($4.8 \times$ to $5.9 \times$). HAC provides optimal balance with aggressive compression ($50 \times$ to $85 \times$) while maintaining competitive quality ($< 0.3$ dB loss vs. Scaffold-GS), validating entropy coding's effectiveness.

Among component-wise methods, CompGS also shows balanced performance with moderate quality loss ($< 1$ dB) and reasonable compression ($15 \times$ to $45 \times$). LightGaussian demonstrates relatively unsatisfactory trade-offs with significant quality degradation ($2 - 4$ dB PSNR loss) despite moderate compression ($12 \times$ to $15 \times$).

Our benchmark shows modern 3DGS compression achieves $10 \times$ to $85 \times $storage reduction with $< 1$ dB quality loss for top methods. HAC provides optimal quality-efficiency balance for practical deployment, though additional encoding/decoding time is required, while Scaffold-GS suits quality-critical applications. Recent Anchor-based approaches outperform component-wise compression methods.

\section{Conclusion}

We present 3DGS-VBench, the first comprehensive VQA dataset for compressed 3DGS, comprising 660 models and video sequences from 11 scenes across 6 algorithms with systematically designed compression parameters. Through subjective evaluation, we obtain MOS scores and validate dataset diversity and reliability.
Our evaluation of 15 IQA/VQA metrics reveals limitations on 3DGS-specific distortions and establishes a standardized benchmark comparing 6 algorithms across storage efficiency and visual quality. This work enables specialized VQA model training and opens possibilities for robust learnable QA metrics and 3DGS compression optimization strategies.


{
    \small
    \bibliographystyle{ieeetr}
    \bibliography{main}
}


\end{document}